\documentclass[twocolumn,showpacs,amsmath,amssymb,aps,prl,nobalancelastpage]{revtex4}
\usepackage{amsthm}
\usepackage{newlfont}
\usepackage{graphicx}
\usepackage{subfigure}
\usepackage{float}
\usepackage[section]{placeins}
\usepackage{psfrag}
\usepackage{caption2}
\usepackage{flafter}
\usepackage{bm}%
\def\fnum@figure{\figurename\thefigure}
\renewcommand{\figurename}{Fig.}

\begin{document}
\title
 {Multistability and Switching in Oppositely Directed Saturable Coupler}
 \author{K. Nithyanandan$^1$, A. K. Shafeeque Ali$^2$, M.P.M. Nishad$^3$, K. Porsezian$^2$, P. Tchofo Dinda$^1$ and P.Grelu$^1$}
 \affiliation{1. Laboratoire Interdisciplinaire Carnot de Bourgogne, CNRS, University of Burgundy-Franche-Comté,
21078 Dijon, France\\2.Department of Physics, Pondicherry University, Pondicherry 605014, India\\ 3. Department of Mathematics, Farook College, Kozhikode, Kerala 673632, India}

\begin{abstract}
We report a novel optical multistability in two core oppositely directed saturable coupler (ODSC) with negative index material (NIM) channel. The dynamics are studied using the Langrangian variational method and Jacobi elliptic functions are used to construct the analytical solutions. The ODSC exhibits a bandgap as a consequence of the effective feedback mechanism due to the opposite directionality of the phase velocity and Poynting vector in the NIM channel. We observe that the system admits multiple stable states for some control parameter, which is a result of the combination of nonlinear saturation and the unique backward-coupling mechanism in the NIM channel. The number of multiple stable states increase with the strength of the nonlinear saturation. Taking the advantage of multiple stable states, one can construct ultrafast switching devices with a flexible switching operation. The studies on the transmission characteristics of the ODSC implies the existence of multiple transmission resonance windows, which could help in the realization of gap soliton in homogeneous systems like coupler. With more degrees of design freedom and controllability, the ODSC could be the choice for a future generation of all-optical switching.
\end{abstract}

\maketitle

\section{Introduction}
The concept of directional couplers proposed in the early 1980's, brought a new paradigm in the ultrafast long haul optical communication system.  Ever since its invention by Jensen  \cite{Jensen:82}, the nonlinear directional coupler (NDC) remains as a fundamental building block of the all-optical communication system. Over the past couple of decades, the NDC earns a considerable interest both at fundamental and applied level \cite{Jensen:82,Wa:85}. The fundamental interest relies on the study of control and manipulation of light by light, while the applied interest is primarily driven by its wide application in all-optical ultrafast switching, signal processing, logic operation \cite{Jensen:82,Wa:85,agrawal_1:08}. The power switching in the NDC is a result of the evanescent coupling of light between the participating waveguides \cite{agrawal_1:08}.  The transmission characteristics of NDC differs based on the design architecture of the coupler, for instance, X junction \cite{Sabini}, cascaded NDC \cite{Aitchison}, bent coupler \cite{Schmidt}, to mention few. In all the cases, the input and output fields are in the same direction and thus the direction of light propagation is preserved, and hence named as directional coupler \cite{agrawal_1:08,Yariv:73,Nithyanandan:13,Venugopal:12}.

\begin{figure}[h]
\begin{center}
\includegraphics*[height=3.5 cm, width = 5 cm]{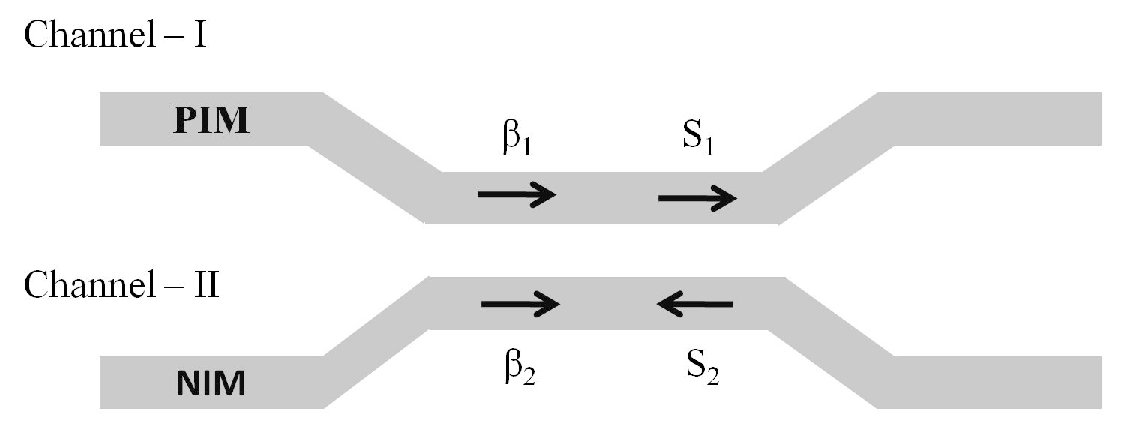}
\caption {\footnotesize{Schematic diagram of oppositely directed coupler with NIM channel.}}
\label{odc}
\end{center}
\end{figure}
\begin{figure*}[htb!]
\begin{center}
 \subfigure[$\gamma_{1,2} =0, \Gamma_{1,2}$=0]{\label{1}\includegraphics[width=4.2 cm]{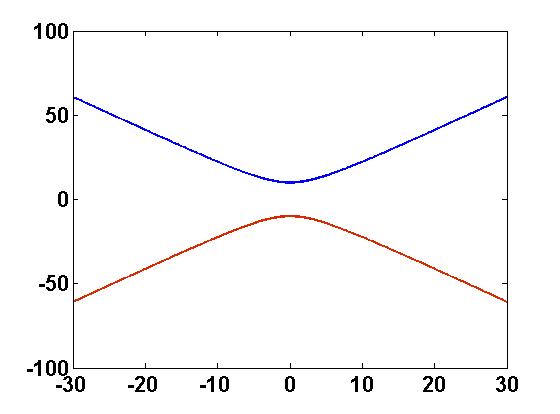}}
 \subfigure[$\gamma_1=\gamma_2=1$]{\label{2}\includegraphics[width=4.2 cm]{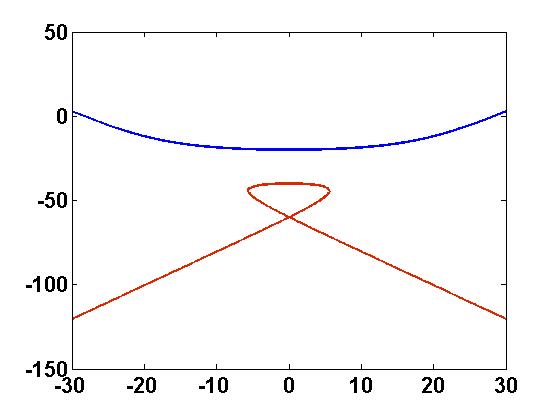}}
  \subfigure[$\gamma_1=1;\gamma_2=0.5$]{\label{3}\includegraphics[width=4.2 cm]{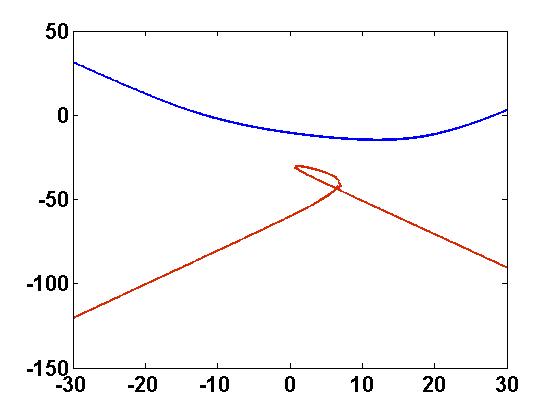}}
 \subfigure[$\gamma_1=0, \gamma_2=0$]{\label{4}\includegraphics[width=4.2 cm]{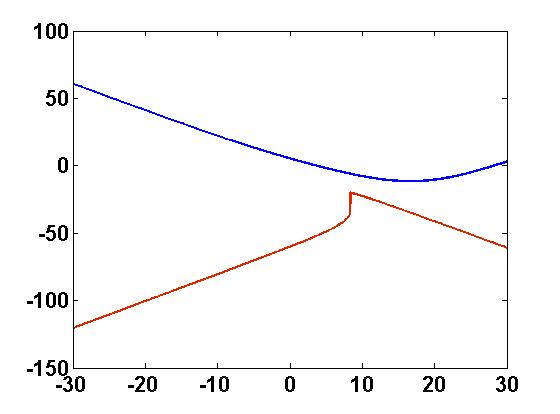}}\\
  \subfigure[$\gamma_{1,2} =1, \Gamma_{1,2}$=1]{\label{5}\includegraphics[width=4.2 cm]{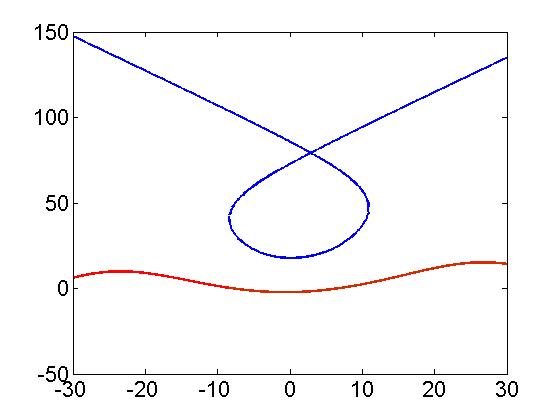}}
 \subfigure[$\gamma_{1,2}=1, \Gamma_1\neq\Gamma_2$]{\label{6}\includegraphics[width=4.2 cm]{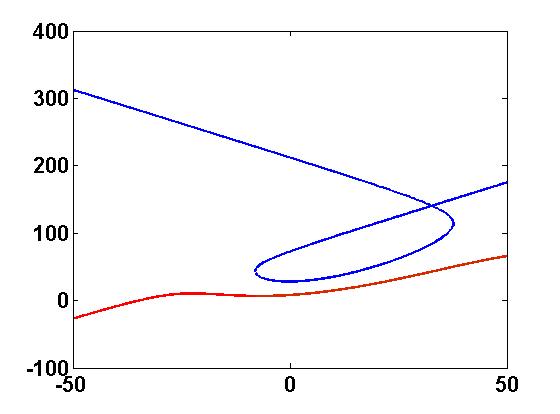}}
   \subfigure[$\gamma_{1}\neq\gamma_2, \Gamma_{1,2}=0.1$]{\label{7}\includegraphics[width=4.2 cm]{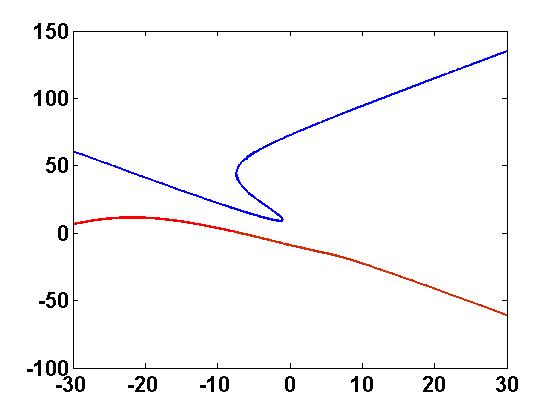}}
 \subfigure[$\gamma_{1}\neq\gamma_2, \Gamma_{1}\neq\Gamma_2$]{\label{8}\includegraphics[width=4.2 cm]{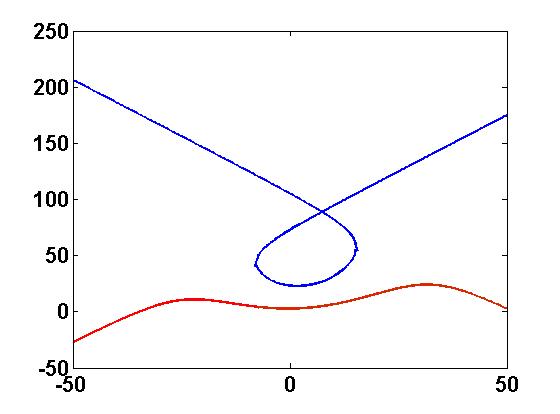}}
\caption {$\delta-q$ curves of ODSC. Top panel corresponds to the Kerr-type nonlinearity, while the bottom panel features PIM-NIM with saturable nonlinearity}
\label{bandgap}
\end{center}
\end{figure*}
One of the most sophisticated advancements in the optical material engineering is the prediction of material showing simultaneous negative permittivity and permeability \cite{Veselago}. The first prediction of such unconventional electromagnetic wave propagation dates back to the late 1960's by Veselago \cite{Veselago}, which remains hypothetical until recently, when Pendry demonstrated an artificially engineered material showing negative  $\epsilon$ and $\mu$, and recognized as Left-handed materials (LHMs) or negative index materials (NIMs) \cite{Pendry_1:00,Smith:00}. Such materials show many exotic and remarkable electromagnetic phenomena like negative refraction, reversed Cerenkov radiation, reversed Doppler shift, reversed Goos-Hanchen shift and reversal of Fermat's principle, which are all found to be uncharacteristic to the naturally available materials~\cite{pendry_1:05,Ramakrishna:05,Caloz:06}. Taking the advantages of the unique properties of NIM, there were many theoretical and experimental results reported at different settings in optics. Among that, incorporation of NIM in one or more coupler channel has been of particular interest, owing to its rich physics.  In such coupler system, the wave vector of the electromagnetic wave is antiparallel to the Poynting vector \cite{pendry_1:05,Ramakrishna:05,Caloz:06} and therefore, the input and output fields are in the opposite direction, and therefore called as the oppositely directed coupler (ODC).  The schematic diagram of ODC with positive index material (PIM) in channel 1 and NIM in channel 2 is depicted in the Fig.\ref{odc}. The performance of ODC is quite similar to the mirror, such that, any forward propagating light in the PIM channel couples continuously with the NIM, where it flows in the backward direction. Due to this unusual property of NIM channel, the ODC shows distinct behavior that is not exhibited by conventional optical couplers~\cite{xiangmod,ShafeequeAli:15,Zhang:15}. One such interesting observation is the existence of the bandgap in ODC, which would other be impossible in the case of conventional couplers.  Litchinitser \emph{et al}, first reported the existence of band gap and bi-stability in ODC, and remarked the bandgap is a result of the intrinsic effective feedback mechanism provided by the NIM channel~\cite{Litchinitser:07}.  Any optically driven system is said to be bistable, if there exist two stable states for the same input power \cite{Gibbs,Bowden}. If they system admits more than two output states, then such phenomenon is known as multistability \cite{Gibbs,Bowden}.  The existence of multiple stable states was predicted theoretically and experimentally accomplished in various optical systems such as Fabry-Perot nonlinear resonators \cite{Winful}, in periodic structures like Bragg Grating \cite{Winful} and also in NDC with external feedback mechanisms \cite{Winful,Vitrant,Nina,Thirstrup,Chevriaux}.

The bistability/multistability is a general phenomenon observed in many domains of optics, which manifest as a result of the integration of nonlinearity and feedback system \cite{Gibbs,Bowden}. The seminal work of Kaplan has opened the interest on the investigation of multi-stable states for the given input control parameters \cite{Kaplan:85}.  The interest in the multistability lies in the fact that, one could construct ultrafast all-optical switches taking the advantages of the multiple stable states \cite{Felber,Erma:86}. Besides being able to perform as a switching, there have been many applications of multi-stable states, for instance, in the development of devices such as memories \cite{Mazurenko,Assanto}, amplifiers \cite{Yosia:07} and so on.

Following the Ref.~\cite{Litchinitser:07}, there have been many interesting works in the direction of exploring the bi-stability in ODC~\cite{coelho:13,Shafeeque:16}. As of now, the reported analysis in ODC has been limited to Kerr law type nonlinearity. However, it has been a well-established fact that the Kerr-law exhibit a behavior, where the refractive index change is proportional to the intensity. This would not be the case for all powers, when the input intensity is sufficiently high, the higher order nonlinear susceptibility inevitably comes into the picture and eventually saturates the nonlinear response, which is termed as saturable nonlinearity (SNL).  Also, in some high nonlinear materials such as glasses, polymers, semiconductor doped fiber system,  the nonlinear is no longer increases indefinitely with intensity, but levels off towards a saturation even at moderate intensity. The effect of saturation in the nonlinear coupler was extensively analyzed by Stegeman and co-workers~\cite{stegeman_1:87} and predicted that the nonlinear saturation significantly influences the transmission and operational characteristics of the coupler. Experimental work on the saturated couplers was carried out by Caglioti and co-workers in Refs.~\cite{caglioti_1:87}. One of the significant outcomes of the investigation was the requirement of low switching power for the saturated coupler, as against the high power threshold for Kerr-type nonlinear couplers \cite{chen:90,begin:94}.

Recently, there were many interesting theoretical results to study the effect of nonlinear saturation in NIM both in the context of conventional fibers and couplers~\cite{xiang:2011,lapine:2014,alves:33,tasting:2012}.  It is informative to know that the negative index materials, so far realized experimentally are made out of a periodic array of split ring resonators (SRRs) and metallic wires \cite{pendry_1:05,Ramakrishna:05,Caloz:06,Shelby}. In the perspective of nonlinear science, there has been a continuing effort to realize LHM associated with high degree of nonlinearity \cite{lapine:2014}. The solution for that is by inserting nonlinear elements in the SRR \cite{pendry_2:99}, or by embedding the SRR in a nonlinear dielectric media~\cite{zharov:03}. One such innovation of producing high nonlinear NIM is by developing a metal structure with materials like GaAs and LiNbO$_3$ \cite{padilla_06}. Those materials show very high nonlinearity which was modeled as NIM encompassed with saturable nonlinearity~\cite{padilla_06,klein:06}. An extensive theoretical analysis of the effect of nonlinear saturation in LHM was reported by Maluckov \emph{et al}, and predicted the existence of bistability region at some parameter space \cite{maluckov:08}.  Thus motivated by the theoretical and experimental realization, we would like to consider a dual-core nonlinear saturated coupler with one core made up of NIM. We study the switching characteristics of the ODC, and report a unique multistable features aided by SNL in NIM, what is the to the best of your knowledge is a first observation in the context of ODC
\section{Theoretical Model}
The coupled mode equation governing the propagation of the modal field $u_{1,2}$ in the saturated ODC is given by the coupled nonlinear schr\"{o}dinger equations as follows (CNLSE) \cite{Litchinitser:07,Shafeeque:16,stegeman_1:87,caglioti_1:87,maluckov:08,alves:33}

\begin{align}
\begin{split}
\label{thm1}
i\,\sigma_j\,\frac{\partial u_j}{\partial z }+\frac{i}{v_{jg}} \frac{\partial u_j}{\partial t }-\frac{\beta_{2j}}{2}\,\frac{\partial^2 u_j}{\partial t^2}+ \sum_{n=1,n\neq j}^{2}k_{j n}\,u_n \,e^{(-1)^j i\delta z} \\ +\gamma_j \frac{f (\Gamma|u_j|^2)}{\Gamma} u_j =0,
\end{split}
\end{align}

\noindent where, $j$ corresponding to 1 and 2 denotes the PIM and NIM channel, respectively. $v_{jg}$ is the group velocity and $\delta=\beta_1-\beta_2$ is difference in the propagation constant.  $\beta_{2j}$, $\gamma_j$ and $k_{jn}$ represent the coefficients of group velocity dispersion, nonlinearity, and coupling, respectively. $\sigma$ indicates the sign of the refractive index based on the relation $n_j = \sqrt{\epsilon(\omega_0) \mu(\omega_0)}$. To account for  NIM, $\sigma$ takes negative sign, while it is positive for PIM channel.  Its worth noting, that the losses in the present case are disregarded for simplicity, as it only expected to increase the required input power level for the desired transmission. The last term in the Eq. (\ref{modeleqn}) accounts for the saturation of the nonlinearity.

As is it evident the conventional Kerr-law nonlinearity exhibits a linear relation between refractive index and the intensity. However, at very high intensities, or material doped with high nonlinear elements show different behavior from the conventional Kerr-law response. The nonlinear response of such materials is no longer monotonously increases with intensity, instead level off to a constant value after a power threshold known as saturation power. Let us assume, the intensity of the light be $I$, then for the Kerr law, the change in the refractive index ($\Delta$n) is proportional to $I$. Whereas, in the case of SNL, the index change happens to be proportional to $I/(1+\Gamma I)$. One can infer, that the system exhibits a Kerr-law behavior at low input intensities, which eventually tends to saturate at higher intensities. Therefore, in modeling the pulse propagation, one must use the saturation function for the nonlinear refractive index as in the Refs.~\cite{NIthyanandan_PRA:2013,Gatz:91,Nithyanandan_EPJ,Kumar:96,Nithyanandan_1:14}.  One way of doing is by replacing the nonlinear refractive index corresponding to the cubic Kerr term with saturating index term.  The other means is to derive the governing equation from Maxwell's equation with the incorporation of appropriate saturation function. As the nonlinear saturation model for the NIM is already available in the literature \cite{maluckov:08}, we adopt such model with suitable modification pertaining to ODC as follows~\cite{NIthyanandan_PRA:2013,Gatz:91,Nithyanandan_EPJ,Kumar:96,Nithyanandan_1:14,maluckov:08}.
\begin{eqnarray}
\label{thm2}
f_j(\Gamma |u_j|^2) =\frac{\Gamma |u_j|^2 }{1+\Gamma |u_j|^2},
\end{eqnarray}
\noindent
where, $\Gamma = 1/P_s$ is the saturation parameter and $P_s$ is the saturation power. In principle, the PIM and NIM channel features distinct propagation parameters, however in order to emphasize on the most important physical effects especially, the nonlinear saturation, we consider $v_{1g}= v_{2g}$, $\beta_{21}= \beta_{22}=\beta$ and $k_{12}= k_{21}=k$.

\noindent
Let us assume the following form of steady state solution for Eq. (\ref{thm1})
\begin{equation}
\label{steady_state}
u_{1,2}=\psi_{1,2}\, \exp (i\, q\, z)\,\exp (\mp i \,\frac{\delta}{2}z),
\end{equation}
In the linear regime, the Eq. (\ref{steady_state}) yields a relation as follows
\begin{equation}
\label{linear_band}
\delta^2=4(k^2+q^2)
\end{equation}
The above equation is an indication of the existence of a bandgap for $|\delta| \leq 2 k$. As it is known, the bandgap is a property associated with a periodic system or distributed feedback structures like fiber Bragg gratings (FBGs) \cite{agrawal_1:08}. Interestingly, the formation of the bandgap in a uniform structure like PIM-NIM coupler is indeed unique, which is attributed to the distinct features associated with the NIM. The nonlinear counterpart of the Eq. (\ref{linear_band}) can be written as
\begin{subequations}
\label{modeleqn}
\begin{eqnarray}
\delta=-\frac{k (1+ f^2)}{f}-\frac{P}{1+f^2}(\gamma_1+ \gamma_2 f^2)+\frac{P^2}{2(1+f^2)^2} \\ \nonumber (\Gamma_2\gamma_2 f^4 -\Gamma_1 \gamma_1),
\end{eqnarray}
\begin{eqnarray}
q=-k(\frac{1- f^2}{2f})-\frac{P}{2(1+f^2)}(\gamma_2 f^2-\gamma_1)+\frac{P^2}{(1+f^2)^2} \\ \nonumber (\Gamma_2\gamma_2 f^4 +\Gamma_1 \gamma_1).
\end{eqnarray}
\end{subequations}
\noindent
where $P=\psi_1^2+\psi_2^2$ is the total power associated with the coupler and $f=\psi_2/\psi_1$ is the ratio of the backward to forward propagating wave amplitude. For better insight, we show in the Fig. \ref{bandgap}, a detailed bandgap analysis with a particular emphasize on the nonlinear saturation. The top panel (Fig. \ref{1}-\ref{4}) shows the $\delta - q$ curves for Kerr nonlinearity and the bottom panel (Fig. \ref{5}-\ref{8}) corresponds to saturable nonlinearity. The existence of bandgap can be readily observed from all the cases illustrated in Fig. \ref{bandgap}. It's worth mentioning that the formation of the bandgap is one of the signatures of a periodic structure such as photonic crystals or distributed feedback structure like the case of fiber Bragg gratings.  But the system considered here is homogenous without any periodicity or feedback mechanism. However, the system still admits bandgap, owing to the effective feedback mechanism supported by the NIM channel. The effective feedback mechanism is a result of the unique backward-coupling as a consequence of the opposite directionality of the phase velocity and Poynting vector. Therefore, the forward propagating light in the PIM channel couples with the NIM, where it flows in the backward direction. The top panel of the Fig. \ref{bandgap} is similar to the results reported in Ref.\cite{Litchinitser:07} for the case of PIM-NIM with Kerr nonlinearity. Fig. \ref{1} and \ref{2} corresponds to the case of linear and nonlinear $\delta - q$ curves and it is similar to the observation of bandgap in the fiber Bragg gratings \cite{Porsezian:05}. The bottom panel corresponds to the case of saturable PIM-NIM coupler, where one can witness the remarkable difference from the Kerr counterpart, which paves the way for the interesting observation delineated in the paper. The qualitative difference between the Kerr and SNL relies on the shape of the $\delta-q$ curve, for instance, the loop is noticed at the upper branch for SNL, while the loop is observed for the lower branch in Kerr-type ODC. Regardless of the nature of the different coefficients, the bandgap persists in all cases, and thus confirm inherent property of the PIM-NIM coupler.\\
\begin{figure*}[ht!]
\begin{center}
 \subfigure[$\Gamma$=0.0]{\label{OB}\includegraphics[height=5.5 cm, width=7.2 cm]{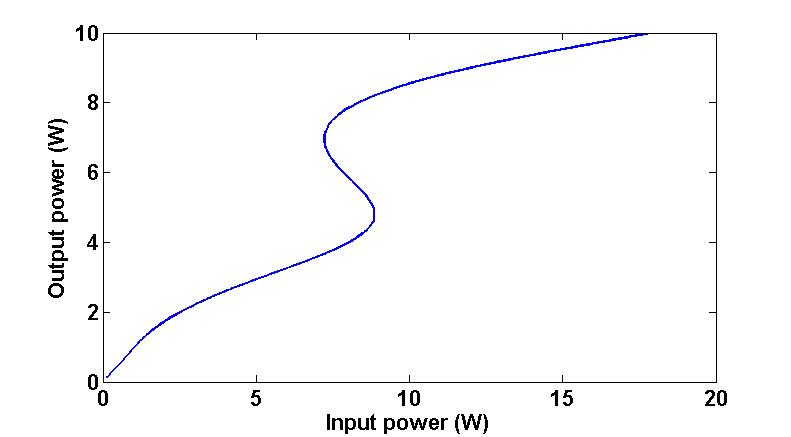}}
 \subfigure[$\Gamma$=0.5]{\label{OM}\includegraphics[height=5.5 cm, width=7.2 cm]{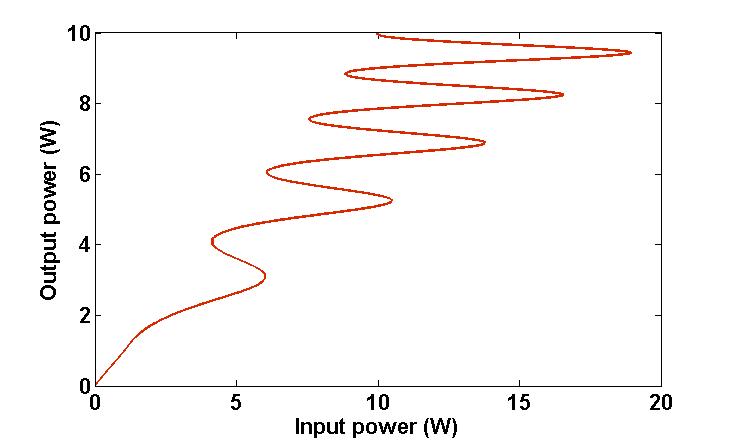}}
\caption {The plot of transfer function of ODSC for (a) Kerr-type nonlinearity, and (b) Saturable nonlinearity.}
\label{OB_OM}
\end{center}
\end{figure*}
Now, we apply a variational approach based on Lagrangian formalism to the propagation model given by Eq. (\ref{thm1}). The seminal work of Anderson in fibers using variational approach has emerged as benchmark analytical tool to explore the underlying physics of nonlinear fiber optics \cite{Anderson}. The Lagrangian of our system can be written as

\begin{equation}
\label{LA}
L=\sum_{j=1}^2 L_j +L_C..
\end{equation}

with
\begin{subequations}
\begin{align}
\begin{split}
\label{LA1}
L_j= i\,\frac{\sigma_j}{2}\,\left(u_j\frac{\partial u_j^*}{\partial z }-u_j^* \frac{\partial u_j}{\partial z }\right)+ i\,\frac{1}{2 v_{jg}}\,\left(u_j\frac{\partial u_j^*}{\partial t }-u_j^* \frac{\partial u_j}{\partial t }\right)\\
-\beta_{2j} \,|\frac{\partial u_j}{\partial t }|^2-\frac{\gamma_j}{2}\,|u_j|^{4}+ \frac{\Gamma_j \gamma_j}{3}|u_j|^{6},
\end{split}
\end{align}
and
\begin{eqnarray}
\label{inter_LC}
L_C= -k(u_1^* \,u_2 \,e^{-i\delta z}+u_2^* \,u_1 \,e^{i\delta z}).
\end{eqnarray}
\end{subequations}
where, $L_j$ is the Lagrangian corresponding to PIM and NIM channel and $L_C$ is the interaction lagrangian. We proceed by assuming a trial function of the Gaussian form as follows,
\begin{equation}
\label{gauss}
u_j(t,z)= \psi_j(z)\, e^{(-\rho \,t^2 + i \,\phi_j(z))}.
\end{equation}
where, $\psi_j(z)$  and  $\phi_j(z)$ is the field amplitude and phase variable, respectively and $\rho$ is a constant. The reduced lagrangian has the form
\begin{equation}
\label{reduce_LA}
\langle L\rangle=\int_0^\infty L dt,
\end{equation}
Substituting Eq. (\ref{LA}) in Eq. (\ref{reduce_LA}), we get the reduced Lagrangian,

 \begin{eqnarray}
\langle L\rangle= -\frac{1}{2}\beta \sqrt{\frac{\pi\rho}{2}}(\psi_1^2+\psi_2^2)
-\frac{1}{8}\sqrt{\frac{\pi}{\rho}} (\gamma_1 \psi_1^4+ \gamma_2 \psi_2^4) \,\,\,\,\,\,\,\,\,\,\,\,\\\nonumber
+ \frac{1}{6} \sqrt{\frac{\pi}{6 \rho}}(\Gamma_1 \gamma_1 \psi_1^6+\Gamma_2 \gamma_2 \psi_2^6)
+\frac{1}{2}\sqrt{\frac{\pi}{2\rho}}(\psi_1^2 \frac{\partial \phi_1}{\partial z} -\psi_2^2\frac{\partial \phi_2}{\partial z})\\ \nonumber
 -\frac{\sqrt{\frac{\pi}{2}}k \psi_1 \psi_2}{2 \sqrt{\rho}} (e^{-i\delta z+\phi_1-\phi_2}+e^{i\delta z+\phi_1-\phi_2}).
\end{eqnarray}

Now varying the effective lagrangian with respect to the variational parameters $\psi_1(z)$, $\psi_2(z)$, $\phi_1(z)$ and $\phi_2(z)$, one can arrive the following set of four coupled differential equations of the form,
\begin{subequations}
\label{modeleqn}
\begin{eqnarray}
\frac{\partial \phi_1}{\partial z }=\beta\rho+ \frac{\gamma_1 \psi_1^2}{\sqrt{2}}-\frac{\Gamma_1 \gamma_1}{\sqrt{3}}  \psi_1^4+\frac{k \psi_2}{\psi_1} \cos(\delta z+ \phi_1-\phi_2),
\end{eqnarray}
\begin{eqnarray}
\frac{\partial \phi_2}{\partial z }=-\beta\rho- \frac{\gamma_2 \psi_2^2}{\sqrt{2}}+\frac{\Gamma_2 \gamma_2}{\sqrt{3}}  \psi_2^4-\frac{k \psi_1}{\psi_2} \cos(\delta z+ \phi_1-\phi_2),
\end{eqnarray}
\begin{eqnarray}
\frac{\partial \psi_1}{\partial z }=k \psi_2 \sin(\delta z+ \phi_1-\phi_2),
\end{eqnarray}
\begin{eqnarray}
\frac{\partial \psi_2}{\partial z }=k \psi_1 \sin(\delta z+ \phi_1-\phi_2).
\end{eqnarray}
\end{subequations}
\normalsize
To proceed further, we define the new phase variable,
\begin{equation}
\phi(z)=\phi_1(z)-\phi_2(z).
\end{equation}
Using the above defined phase variable above sets of four coupled differential equations reduces to three as follows
\begin{subequations}
\label{modeleqn}
\begin{eqnarray}
\frac{\partial \phi}{\partial z }=2\beta\rho+ \frac{1 }{\sqrt{2}}(\gamma_1 \psi_1^2+\gamma_2 \psi_2^2)-\frac{1}{\sqrt{3}}  ( \Gamma_1 \gamma_1 \psi_1^4+ \\ \nonumber \Gamma_2 \gamma_2 \psi_2^4)+k (\frac{\psi_2}{\psi_1}+\frac{\psi_1}{\psi_2}) \cos(\delta z+ \phi),
\end{eqnarray}
\begin{eqnarray}
\frac{\partial \psi_1}{\partial z }=k \psi_2 \sin(\delta z+ \phi),
\end{eqnarray}
\begin{eqnarray}
\frac{\partial \psi_2}{\partial z }=k \psi_1 \sin(\delta z+ \phi).
\end{eqnarray}
\end{subequations}
The constants of motion of above set of equations are given by
\begin{subequations}
\label{modeleqn}
\begin{eqnarray}
F=\psi_1^2-\psi_2^2=P_1-P_2,
\end{eqnarray}
\begin{eqnarray}
G=4\psi_1\psi_2\cos(\delta z+\phi)+\frac{2}{k}\psi_1^2(\beta\rho+\frac{\delta}{2}+\frac{\gamma_1}{2\sqrt{2}}\psi_1^2)\,\,\,\,\,\,\,
\\ \nonumber +\frac{2}{k}\psi_2^2(\beta\rho+\frac{\delta}{2}+\frac{\gamma_2}{2\sqrt{2}}\psi_2^2)-\frac{2}{3 \sqrt{3} k}(\gamma_1 \Gamma_1 \psi_1^6+\gamma_2 \Gamma_2 \psi_2^6).
\end{eqnarray}
\end{subequations}
If $\delta=0$ and $G=0$, eqs.(9(b)) can be written as
\begin{widetext}
\begin{eqnarray}
\label{power}
\left(\frac{d P_1}{d z }\right)^2= F^2(\beta\rho(\frac{c}{\sqrt{2}} +\frac{2}{3\sqrt{3}} cd -\beta\rho)-c^2(\frac{1}{3\sqrt{6}}d+ \frac{1}{27}d^2+\frac{1}{8}))+F (P_1(\frac{2}{9}c^2d^2+\frac{5}{3\sqrt{6}}c^2 d+\frac{1}{2}c^2-4k^2) \\ \nonumber -\beta\rho(\frac{10}{3\sqrt{3}}c d +2 \sqrt{2}c-4\beta\rho)) +P_1^2(\frac{\beta\rho}{\sqrt{2}}(a+5c+2\sqrt{6}cd- 4\sqrt{2}\beta\rho)+\frac{c^2}{3}(5\sqrt{\frac{2}{3}}d-\frac{5}{3}d^2+\frac{ad}{\sqrt{6}c}
+\frac{3a}{4c}-12\frac{k^2}{c^2}+\frac{9}{4}))\\ \nonumber+P_1^3 \gamma_2 (\frac{2}{27}abd + \frac{a}{\sqrt{6}}(d+\frac{b}{3})+\frac{1}{2}(a+c)+\frac{5}{3}\sqrt{\frac{2}{3}}dc+\frac{20}{27}d^2 c-\sqrt{2}\beta\rho(\frac{\sqrt{2}b\gamma_1}{3\sqrt{3}\gamma_2}+\frac{14d}{3}\sqrt{6}+\frac{\gamma_1}{\gamma_2}+1))\\ \nonumber -P_1^4 \gamma_1\gamma_2 (\frac{4\beta\rho}{3\sqrt{3}}(\frac{\Gamma_1}{\gamma_2} +\frac{\Gamma_2}{\gamma_1})+\frac{\gamma_2}{\gamma_1}(\frac{5d}{3\sqrt{6}}+\frac{5d^2}{9}+\frac{\gamma_1^2}{\gamma_2^28} +\frac{1}{8})+\frac{1}{3}\sqrt{\frac{2}{3}}b+\frac{2}{9}bd+\frac{d}{\sqrt{6}}+\frac{1}{4}) +\frac{P_1^5\gamma_1\gamma_2 \Gamma_2}{3\sqrt{6}}\\ \nonumber (\frac{\gamma_1\Gamma_1}{\gamma_2\Gamma_2}+\frac{\Gamma_1}{\Gamma_2}+\frac{\gamma_2}{\gamma_1} (\frac{2\sqrt{6}}{3}d+1)+\frac{2\sqrt{6}}{3}b+1)-\frac{P_1^6\gamma_1\Gamma_1}{27}(\gamma_1\Gamma_1+2\gamma_2\Gamma_2 +\frac{\gamma_2^2 \Gamma_2^2}{\gamma_1\Gamma_1}).
\end{eqnarray}
\end{widetext}

Where $a=F\gamma_1$, $b=F\Gamma_1$, $c=F\gamma_2$ and $d=F\Gamma_2$. The periodic solution of Eq.(\ref{power}) can be obtained by using Jacobian elliptic function. We consider the solution of Eq.(\ref{power}) takes the form
\begin{equation}
\label{sol}
P_1(z)=A\,cn[\Omega(z-L),m].
\end{equation}
where $A$ and $\Omega$ are constants and $m$ is the modulus of elliptic function. Substituting Eq.(\ref{sol}) in Eq.(\ref{power}) and equating the coefficients of cn function, one can find the values of constants as given below,
\begin{subequations}
\label{modeleqn}
\begin{eqnarray}
A=\frac{m\Omega}{\sqrt{\gamma_1\gamma_2(\frac{\gamma_2}{\gamma_1}a_1-\beta\rho a_2+ a_3)}},
\end{eqnarray}
\begin{eqnarray}
\Omega=\frac{F}{A (m^2-1)} \sqrt{c^2(a_4-\beta\rho a_5)}
\end{eqnarray}
\begin{eqnarray}
a_1=d(\frac{5}{3\sqrt{6}}+d)+\frac{1}{8}(\frac{\gamma_1^2}{\gamma_1^2}+1),
\end{eqnarray}
\begin{eqnarray}
a_2=\frac{1}{3\sqrt{3}}(4\frac{\Gamma_1}{\gamma_2}+\frac{\Gamma_2}{\gamma_1}),
\end{eqnarray}
\begin{eqnarray}
a_3= \frac{1}{3}\sqrt{\frac{2}{3}}b+\frac{d}{\sqrt{6}}+\frac{2}bd
\end{eqnarray}
\begin{eqnarray}
a_4= \frac{1}{3\sqrt{6}}d+\frac{1}{27}d^2+\frac{1}{8}
\end{eqnarray}
\begin{eqnarray}
a_5= \frac{c}{\sqrt{2}}-\frac{2}{3\sqrt{3}}cd+1
\end{eqnarray}
\end{subequations}
If we assume to excite the channel 1, \emph{i,e},  $P_1(0)=P_0$, $P_2(L)=0$ then $F=P_1^2(L)$, where $P_j(0)$ and $P_j(L)$, indicate the values of power at $z=0$ and $z=L$, respectively. Then, the calculated transfer function can be given as
\begin{equation}
\frac{P_1(L)}{P_1(0)}=\frac{1}{cn[\Omega L,m]}.
\label{TC}
\end{equation}
\noindent
\begin{figure}[ht!]
\begin{center}
\includegraphics*[height=5.5cm, width = 7cm]{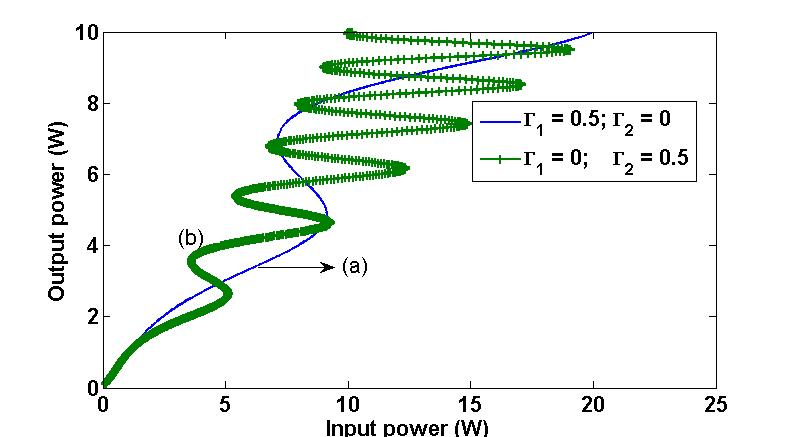}
\caption {The plot of transfer function with either one of the channel is saturated.}
\label{G1_G2}
\end{center}
\end{figure}
\section{Results and Discussion}
The Eq. (\ref{TC}) gives an estimate of the existence of the stable states in the system as a function of power. The plot of the variation of output versus input power could provide many interesting information, and one of the feature with wide interest is the bi- or multi- stability.

\begin{figure}[ht!]
\begin{center}
\includegraphics*[height=5.5cm, width = 7cm]{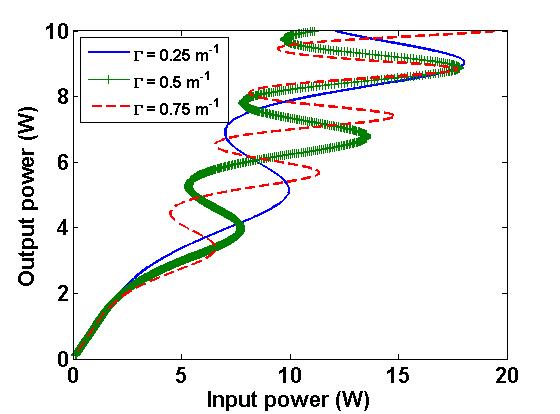}
\caption {The plot of transfer function showing multistability for certain set of saturation parameter.}
\label{Transfer_function}
\end{center}
\end{figure}
\begin{figure}[ht!]
\begin{center}
\includegraphics*[height=5.5cm, width = 7cm]{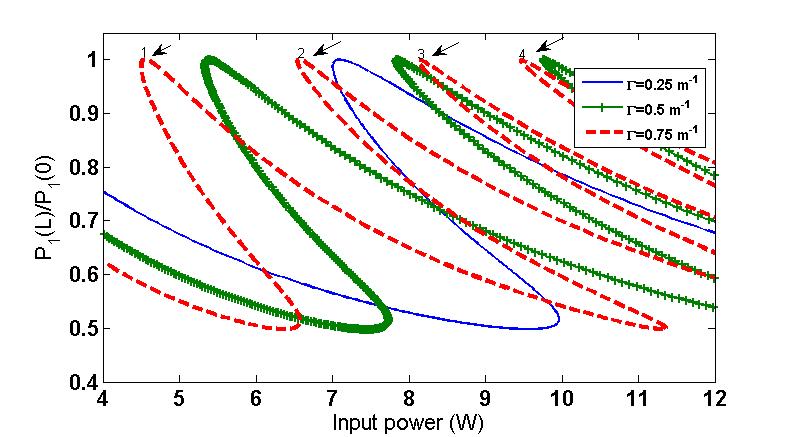}
\caption {Plot showing the transmission coefficient for different values of saturation parameter.}
\label{Transmission_1}
\end{center}
\end{figure}
Multistability is a phenomenon where the system exhibit more than two stable states for a certain control parameter like power. Although some extensive results for the bi-stability in the two core PIM-NIM coupler can be found elsewhere, including our recent work, we will, nevertheless,  reproduce few of the interesting results for demonstration purpose.  The typical parameters of choice in our entire analysis are as follows: $\gamma =0.5$ W$^{-1}$ cm$^{-1}$, $\beta=0.62$ps$^2$ cm$^{-1}$, $m=0.3$, and $k=0.38$ cm$^{-1}$. Fig. \ref{OB_OM} delineate the variation of output power versus input power for the case of Kerr-type nonlinearity, which can be achieved by setting $\Gamma =0$. It is apparent from the Fig. \ref{OB}, that the system inevitably admits bi-stability, regardless of the presence of SNL, which confirms the previously published results.  This attribution is a consequence of the effective feedback mechanism caused by the opposite directionality of the phase velocity and Poynting vector, as observed in Ref.~\cite{Litchinitser:07,Shafeeque:16} Fig. \ref{OM} depicts the effect of nonlinear saturation in the stable states as a function of input power. One can straightforwardly notice that the SNL brings more stable states for the same power. The plot of the transfer function shown in the Fig.~\ref{OM} can be interpreted as the co-existence of bi- and multi-stability for the same setting. It can be inferred that the multi-stability observed in the Fig.~\ref{OM} builds initially from the bi-stable states for low input powers, and with raise in the input power the system brings more stable states. This is in direct correspondence to our earlier discussion on SNL, such that, the system behaves like Kerr-law media for low input power which accounts for bistability as observed in Ref. \cite{Litchinitser:07,Shafeeque:16}. As the input power increases above the saturation threshold, then the system is said to be in the saturation regime and support more stable states.\\

\begin{figure}[ht!]
\begin{center}
\includegraphics*[height=5.5cm, width = 7cm]{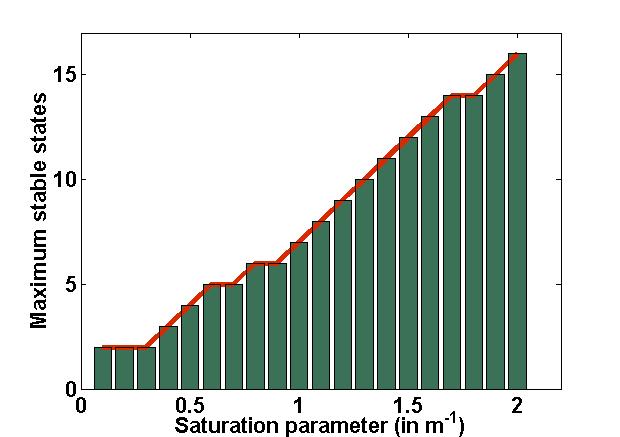}
\caption {Diagram shows the monotonous increase of the stable states with $\Gamma$.}
\label{states_Gamma}
\end{center}
\end{figure}

The occurrence of the multi-stability in ODSC is identified as a result of the combination of nonlinear saturation and the opposite directionality of phase velocity and Poynting vector of NIM channel. It's worth mentioning that the transfer function supported by the nonlinear saturation function described in Eq. (\ref{thm2}) can inherently support bi-stability, see Ref.\cite{Gatz:91,Nithyanandan_EPJ,Kumar:96,Nithyanandan_1:14,maluckov:08}. Taking advantage of the bi-stability supported by saturation function, the distributed effective feedback mechanism caused by NIM channel can contribute in the same direction, enabling more stable states resulting in multi-stability.   For further insight into the effect of nonlinear saturation, we depict in the Fig. \ref{G1_G2}, the plot of the transfer function with either one of the channel as saturated. Curve (a) represents the case where only PIM channel is saturated $\Gamma_1\neq0, \Gamma_2=0$, and curve (b), corresponds to the case of saturated NIM channel $\Gamma_1=0, \Gamma_2\neq0$.

It is apparent from curve (a) of  Fig. \ref{G1_G2}, when the saturable coefficient of the NIM channel is zero, there exist only two stable states, which concur with the results of Ref.~\cite{Litchinitser:07}. Thus, one can infer, it is only the saturation of nonlinearity in the NIM channel (see curve b) is critical to the existence of higher order multi-stable states, which is a clear manifestation of our earlier interpretation on the multi-stability in ODSC.

\begin{figure}[ht!]
\begin{center}
\includegraphics*[height=5.5cm, width = 7cm]{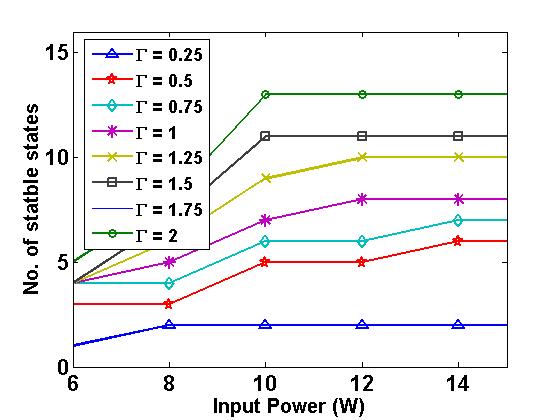}
\caption{Plot showing the variation of number of stable states as a function of input power.}
\label{srtates_power}
\end{center}
\end{figure}

To demonstrate the impact of the strength of the nonlinear saturation, we depict in the Fig. \ref{Transfer_function}, the transfer function for some representative values of saturation parameter. It is obvious to note that more stable states are generated with the increase in the strength of the nonlinear saturation. It is a well-established fact that the phenomenon OB is in close proximity with the formation of gap solitons, as evidenced from different studies ranging from lattice to photonic crystal fibers \cite{Lidorikis:04,Chen:87}. Gap soliton is typically a signature of the periodic structure exhibiting band gap \cite{Lidorikis:04,Chen:87,Christodoulides:89}, however, in the present case of the coupler, we show that formation of gap soliton is a consequence of distributed feedback mechanism supported by NIM channel.  To give some insight into the perspective of gap soliton, we plot in Fig.~\ref{Transmission_1} the transmission coefficient as a function of power. One can notice that the transmission coefficient is equal to unity, which implies the existence of transmission resonances. At this resonance condition, the light that is coupled to the system can support soliton-like static entity, known us the so-called gap soliton.  Unlike the previously reported results, the present system supports multiple transmission resonances window, which could be of potential use for the realization of gap soliton. This possibility of gap soliton in a homogeneous system like coupler makes ODSC as a rich field of scientific curiosity. A detailed study of the gap soliton in saturable PIM-NIM will be addressed in our future communications.\\

For a numerical appreciation and to quantify our earlier interpretation on the influence of nonlinear saturation in the transfer function. We show in the Fig.~\ref{states_Gamma}, the maximum stable states supported by the system as a function of saturation parameter. It is obvious that the number of stable states increases monotonously with the strength of the saturation parameter, which concurs with our early description. Fig.~\ref{srtates_power}, depicts the possible stable states supported by the system as a function of input power for some representative values of $\Gamma$. One can straightforwardly notice that the range of the stable states depends on the input power. For a comprehensive picture, we have also studied the effect of system parameters in the threshold for OB and OM. We identified that the threshold power decreases with increase in nonlinear, dispersion and coupling coefficients. This is quite similar to previously published results \cite{Litchinitser:07,Shafeeque:16} and hence an extensive illustration is needless here.

\section{Conclusion}
In summary, we investigated the optical multistability and switching in a saturable oppositely directed coupler with negative index material channel. The dynamical study has been carried out by using the Lagrangian variational method, and the Jacobi elliptic functions are used to construct the analytical solutions. Unlike the conventional PIM-NIM coupler, the ODSC yields multiple stable states for the same input power. This is attributed to the combination of nonlinear saturation and the unique backward coupling as a result of the opposite directionality of phase velocity and Poynting vector of NIM. We have also identified that the number of stable states increases with the strength of saturation. Further insight implies, only the saturation of NIM is fundamental to the existence of multiple states in the parametric space of interest. Taking the advantage of multiple stable states, one can construct ultrafast switching devices with a flexible switching operation. Also, ODCs offers more degrees of design freedom to maneuver the key features of multistability like the threshold power, hysteresis loop etc., by merely tuning the system parameters.  The studies on the switching characteristic reveal that the ODSC offers many windows of transmission resonance, which could enable one to realize gap solitons in homogenous fiber system like couplers. With all aforementioned results, the proposed ODSC could be a potential prospect for the future generation of all-optical switching.  As there is growing interest in nonlinear metamaterials, researchers are already successful in producing NIM showing saturable nonlinearity. With many new innovations in artificial materials are on its way, we believe our theoretical results can engender many new experiments and could give some guidelines in the design and development of coupler based on negative index materials for various applications.
\section{ACKNOWLEDGMENTS}
 K.P thanks DST, CSIR, NBHM, IFCPAR and DST-FCT Government of India, for the financial support through major projects. P.G. and P.T.D acknowledge the support from CEFIPRA (project 5104-2), LABEX Action and Région Bourgogne. K.N thanks CNRS for post doctoral fellowship.
\bibliographystyle{amsplain}

\end{document}